**Title:** Robust synchrotron-based deep learning algorithm for intracochlear segmentation in clinical scans: development and international validation

**Authors:**

Ashley Micuda[1], Daniel Newsted[2], Nastaran Shakourifar[3], Sachin Pandey[4], Asma Alahmadi[2], Kevin D. Brown[5], Abdulrahman Hagr[6], Jacob B. Hunter[7], Joachim Müller[8], Kristen Rak[9], Hanif M. Ladak[1,2,3,10]*, Sumit K. Agrawal[1,2,3,10]*

*Co-senior authors

S.K.A. is the corresponding author.

**Affiliations:**

[1]Department of Medical Biophysics, Western University, London, Ontario, Canada

[2]Department of Otolaryngology–Head and Neck Surgery, Western University, London, Ontario, Canada

[3]School of Biomedical Engineering, Western University, London, Ontario, Canada

[4]Department of Medical Imaging, Western University, London, Ontario, Canada

[5]Department of Otolaryngology–Head and Neck Surgery, University of North Carolina, Chapel Hill, North Carolina, U.S.A.

[6]King Abdullah Ear Specialist Center (KAESC), King Saud University Medical City, King Saud University, Riyadh, Saudi Arabia

[7]Department of Otolaryngology–Head and Neck Surgery, Thomas Jefferson University, Philadelphia, Pennsylvania, U.S.A.

[8]Department of Otorhinolaryngology–Head and Neck Surgery, Ludwig-Maximilians-University Munich, Munich, Germany

[9]Department of Otorhinolaryngology–Head and Neck Surgery, Comprehensive Hearing Center, University of Würzburg, Würzburg, Germany

[10]Department of Electrical and Computer Engineering, Western University, London, Ontario, Canada

## Abstract

Clinical imaging is routinely used for cochlear implant surgical planning yet lacks the resolution and contrast necessary to visualize the fine intracochlear structures critical for individualized intervention. To address this limitation, an ensemble deep learning model was developed to automatically segment cochlear micro-anatomy from standard clinical scans. The model was trained and validated using an independent internal dataset comprised of paired synchrotron and clinical scans of the same cochlea across various acquisition protocols. Performance was evaluated quantitatively on an unseen internal test dataset and a multi-institutional external test dataset. The deep learning model achieved accurate segmentation of intracochlear anatomy across all tested modalities, outperformed all previously published models, and demonstrated strong viability on the multi-institutional external dataset. Furthermore, anatomical measurements on the automatic segmentations closely matched those obtained from high-resolution ground truth segmentations, confirming reliable estimation of clinically relevant metrics. By bridging the gap between high-resolution imaging and routine clinical imaging, this work provides a practical solution for patient-specific cochlear implant surgical planning and postoperative assessment, advancing the goals of atraumatic insertions and more effective hearing restoration.

## Introduction

Approximately half a billion individuals worldwide live with disabling hearing loss[1] and prevalence is projected to exceed 900 million by 2050[2]. Cochlear implants (CIs) are the standard intervention for restoring sound perception in individuals with severe-to-profound sensorineural hearing loss[3], with over one million devices implanted globally[4]. CIs are surgically implanted neural-prosthetic devices comprised of both external and internal components. A key internal component is the electrode array, which is surgically inserted into the scala tympani (ST), one of three intracochlear ducts, to electrically stimulate the auditory nerve fibers, enabling sound perception.

Despite the success of CIs, hearing outcomes are highly variable across individuals and are influenced by patient-specific factors including anatomical variability[5–8]. Clinical imaging modalities such as helical computed-tomography (CT) and cone-beam CT (CBCT) are routinely used for pre-operative planning[9], with post-operative imaging also becoming more prevalent for patient-specific programming. However, the limited spatial resolution and contrast inherent to these modalities prevents visualization of the fine intracochlear structures, including the ST. Intracochlear morphological variations have important implications for selection of CI electrode arrays[10,11]; residual hearing and structure preservation[10,12,13]; post-operative speech perception[14,15]; and anatomy-based frequency-to-place post-operative tuning[16–18]. Accordingly, the ability to visualize the intracochlear ducts both pre- and post-operatively would provide significant clinical benefit for clinicians and CI users.

One strategy to enable intracochlear visualization is through image segmentation, the process of labelling regions in an image corresponding to particular anatomy. Since intracochlear anatomy is generally not well visualized in clinical resolution imaging, manual segmentation of the intracochlear anatomy is infeasible. To address this challenge, mathematical models have been proposed to automatically infer anatomic locations and perform segmentations in clinical images by training on paired high-resolution and clinical resolution datasets. Previous models have been developed by obtaining three-dimensional (3D) surfaces from image intensity profiles[19], using atlas-based modelling[20,21], statistical shape modelling[22,23], and deep learning[24–30] for various temporal bone structures.

Noble *et al.*[22] pioneered scalar segmentation in the field of CIs with active shape models using paired pre-operative micro-CT (μCT) and clinical CT scans of six cochleae, more recently expanding this model to include an importance weighting and an increased sample size of 16 cochleae[23]. This model has been subsequently applied to pre-operative scans for automatic segmentation and aligned with post-operative scans to assess trauma with short electrodes (less than 25 mm)[22,31,32]. More recently, it has guided pre-operative planning for slim precurved arrays in patients for insertion depths up to 542°[33]. Building on this foundation, Margeta *et al*.[25] introduced *Nautilus*, a web-based research platform that employs a Bayesian joint inference model to constrain a 3D U-Net, a type of convolutional neural network, for automatic scalar segmentation, similarly using μCT as ground truth data. While the model achieved high accuracy in the basal turn, validated recently with histology[34], performance decreased beyond the basal turn (with no metrics reported past 522°).

Despite these advances in clinical scalar segmentation, several limitations remain. First, existing models do not achieve high accuracy in the apical region of the cochlea. Given that the longest commercially available electrodes typically reach 667.1° ± 36.5° in patients[35], with some studies reporting insertion depths beyond 720°[17,36], this restricts the generalizability of current models for longer electrode arrays. Long electrode arrays are capable of achieving high angular insertion depths (AIDs), which has been reported to correlate with improved post-operative speech test scores[37–44], highlighting the need for accurate apical segmentation. Second, all previous models were developed using ground truth datasets derived from μCT, which provides limited soft-tissue resolution and can suffer from blooming artefacts[45]. This constraint may compromise segmentation accuracy and inherently restricts performance to the resolution of the ground truth data. Third, the datasets used for model development were relatively small, likely not capturing the variability seen in the population[5,6,46,47] and risks over-constraining the models. Finally, most approaches were tailored to a single clinical scan type, limiting their applicability across centers that use different imaging modalities (CBCT vs Helical CT), resolutions, protocols, and manufacturers.

To address these limitations, an ensemble deep learning network was developed for automatic scalar segmentation extending to the cochlear apex. Leveraging synchrotron radiation phase-contrast imaging (SR-PCI) as the ground truth dataset allowed for detailed visualization of the cochlear micro-anatomy through the entire length of the cochlea. SR-PCI overcomes the soft-tissue limitations associated with μCT, enabling more accurate ground truth segmentation[7,48–54]. The SR-PCI segmentations were paired with clinical-resolution scans of varying modalities, resolutions, and manufacturers. Various augmentations were then applied, resulting in a dataset comprising 108 unique cochleae and 4,218 scans. The objective of this study was to develop the first scalar segmentation model using SR-PCI ground truth data, which is capable of generalizing across multi-resolution, multi-modality, and multi-institutional clinical scans. The proposed deep learning segmentation model may be integrated into clinical workflows to facilitate greater precision in patient-specific CI surgical planning and post-operative management, ultimately improving patient outcomes.

## Results

### *Automatic Segmentation Model*

A deep learning model was developed to automatically segment the ST and scala vestibuli (SV) in clinical-resolution CT data. The segmentation model was built using the 3D SegResNet ensemble model implemented in MONAI[55]. Internal clinical-resolution CT scans were used as training inputs and ground truth SR-PCI segmentations of the ST and SV were used as training targets. Prior to training, a statistical summary of the training dataset including image intensity, shape, and spacing was computed to guide pre-processing parameter selection. The default algorithm transforms were used which included intensity normalization, random rotation, and random intensity shifting. A five-fold split of the training data was generated using a random seed, with the requirement that left and right cochleae of the same cadaver be in the same fold. An ensemble model was established using segmentation networks trained using each of the five folds, with test predictions formed using majority voting across the individual network outputs. Of the 108 cochleae used in this model, 93 were used for training and validation in the five-fold development (a total of 4,016 scans), and the remaining 15 were used as an unseen test set (94 scans).

### *Global Objective Metrics Evaluation*

Model performance was assessed on 94 unseen test images from the internal dataset acquired using various modalities, resolutions, and protocols on 15 cochleae. Representative examples of the best and worst automatic segmentations of the ST and SV on 300 µm CBCT scans according to DSC scores are illustrated in Figure 1. Close agreement between the ground truth and predicted segmentations can be seen through visual inspection for both examples given. Slight underestimation can be observed laterally for the worst predicted segmentation, shown in Figure 1c and d, but general shape along the length of the cochlea agrees closely with the ground truth data. Clear delineation between the scala and the spiral ligament is evident along the length of the cochlea when overlayed on the SR-PCI scans.

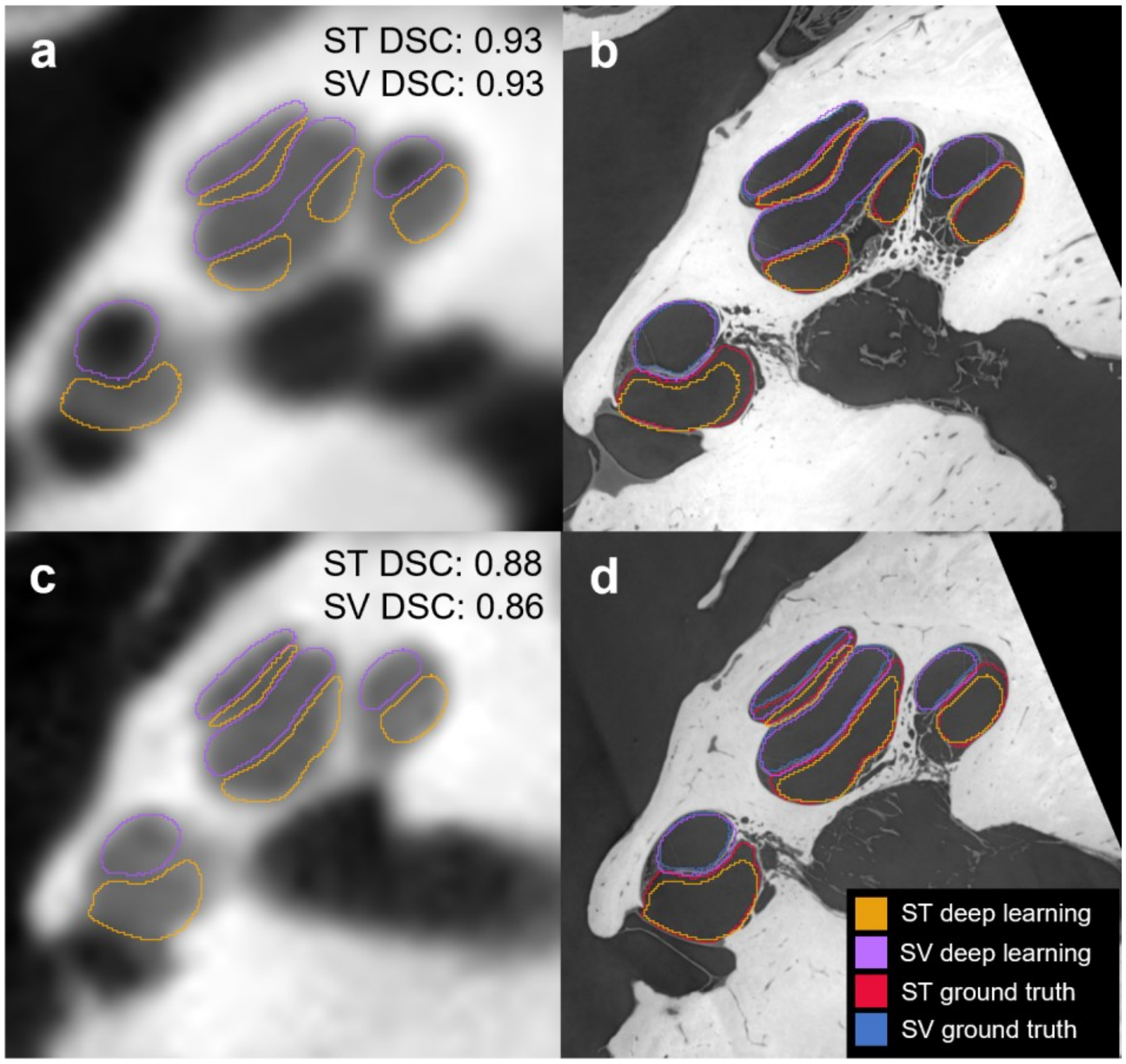

**Fig 1. Automatic segmentation on CBCT compared with ground truth SR-PCI segmentation. (a)** Example of the best automatic segmentation on an unseen 300 µm CBCT test scan, as determined using the dice coefficient score (DSC), and **(b)** the automatic and ground truth segmentation overlayed on a 2D axial cross-sectional SR-PCI slice of the same sample. **(c)** Example of the worst automatic segmentation on an unseen 300 µm clinical CBCT test scan, as determined using the DSC, and **(d)** the automatic and ground truth segmentation overlayed on a 2D axial cross-sectional SR-PCI slice. The ground truth scala tympani (red) and scala vestibuli (blue) overlayed; and the automatic segmentation of the scala tympani (orange), and scala vestibuli (purple) overlayed. The achieved DSC is shown for both scalae in the top right corner.

Segmentation accuracy of the deep learning model was assessed using standard objective metrics, including the Dice similarity coefficient (DSC), max Hausdorff distance (HD), and average HD for the ST and SV, compared to the reference ground truth segmentation derived from SR-PCI data. Across all clinical-resolution scans, for the ST and SV respectively, the model achieved a mean DSC of 0.895 and 0.891, a max HD of 0.392 mm and 0.475 mm, and an average HD of 0.008 mm and 0.008 mm.

Detailed performance by scan type and structure is summarized and compared against previously reported segmentation methods in Table 1. Individual sample scores across scan types are presented as supplements (Supplementary Fig. 1). The proposed approach consistently outperformed previously published techniques across all modalities, particularly for the 300 µm CBCT, which is used most commonly in the literature. The DSC values exceeded 0.90 for both the ST and SV using the proposed model, outperforming all previous reports.

**Table 1. Automatic segmentation model comparison across image modalities and objective metrics.** A single value for the resolution indicates isotropy and three values indicate anisotropy; '-' indicates this data was not reported; n indicates number of cochleae evaluated.

| Literature | Method | n | Modality/ Scanner Type | Dice Similarity Coefficient | | Max Hausdorff Distance (mm) | | Average Hausdorff Distance (mm) | |
|---|---|---|---|---|---|---|---|---|---|
| | | | | ST | SV | ST | SV | ST | SV |
| Current work | Deep learning ensemble SegResNet | 15 | Xoran CBCT (0.1 mm) | 0.908 ± 0.017 | 0.908 ± 0.019 | 0.371 ± 0.092 | 0.407 ± 0.268 | 0.006 ± 0.002 | 0.006 ± 0.002 |
| | | 15 | Xoran CBCT (0.3 mm) | 0.907 ± 0.018 | 0.906 ± 0.020 | 0.376 ± 0.085 | 0.409 ± 0.268 | 0.006 ± 0.002 | 0.007 ± 0.002 |
| | | 14 | Siemens CT, Sharp Protocol (0.350 mm) | 0.898 ± 0.019 | 0.893 ± 0.022 | 0.381 ± 0.079 | 0.391 ± 0.234 | 0.007 ± 0.002 | 0.008 ± 0.002 |
| | | 14 | Siemens CT (0.500 mm) | 0.880 ± 0.019 | 0.880 ± 0.028 | 0.409 ± 0.098 | 0.420 ± 0.240 | 0.009 ± 0.002 | 0.009 ± 0.003 |
| | | 14 | Canon Helical CT, Temporal Bone Protocol (0.591 x 0.591 x 0.5 – 0.789 x 0.789 x 0.5 mm) | 0.885 ± 0.031 | 0.878 ± 0.036 | 0.441 ± 0.116 | 0.448 ± 0.293 | 0.009 ± 0.004 | 0.010 ± 0.004 |
| | | 14 | Canon Helical CT, Soft Tissue Protocol (0.591 x 0.591 x 0.5 – 0.789 x 0.789 x 0.5 mm) | 0.889 ± 0.025 | 0.889 ± 0.025 | 0.418 ± 0.123 | 0.464 ± 0.276 | 0.009 ± 0.003 | 0.009 ± 0.004 |
| | | 4 | GE Helical CT, Temporal Bone Protocol (0.488x0.488x0.625 mm) | 0.898 ± 0.015 | 0.896 ± 0.013 | 0.380 ± 0.123 | 0.604 ± 0.383 | 0.007 ± 0.002 | 0.008 ± 0.001 |
| | | 4 | GE Helical CT, Soft Tissue Protocol (0.488x0.488x0.625 mm) | 0.895 ± 0.014 | 0.895 ± 0.014 | 0.365 ± 0.151 | 0.617 ± 0.416 | 0.008 ± 0.002 | 0.009 ± 0.002 |
| Banalagay *et al.* (2023)[23] | Weighted active shape model | 15 | CT (0.15 × 0.15 × 0.2 mm – 0.4 mm) | 0.83 | 0.83 | - | - | - | - |
| Margeta *et al.* (2022)[25] | Deep learning constrained with Bayesian inference | 23 | CT (0.3 mm) | 0.65 | 0.60 | - | - | - | - |
| Powell *et al.* (2021)[21] | Atlas-based | 8 | CBCT (0.25 mm) | 0.77 ± 0.06 | 0.74 ± 0.05 | 0.78 ± 0.08 | 0.76 ± 0.13 | 0.11 ± 0.01 | 0.12 ± 0.01 |
| Zhang *et al.* (2019)[29] | Deep learning 3D CNN U-net | 11 | CT (0.3 mm) | 0.87 ± 0.016 | 0.86 ± 0.017 | - | - | 0.077 ± 0.016 | 0.084 ± 0.017 |
| Noble *et al.* (2011)[22] | Active shape model | 5 | CT (0.3 mm) | 0.77 ± 0.03 | 0.72 ± 0.04 | 0.66 ± 0.16 | 0.88 ± 0.39 | 0.20 ± 0.02 | 0.21 ± 0.02 |

*Local Objective Metrics Evaluation*

Boundary accuracy was further assessed using Euclidean distance maps for the ST and SV on the 300 µm scans, as shown in Figure 2. Distances between the predicted and ground truth surfaces are shown within the range of -0.3 mm to 0.3 mm, with red indicating overestimation, blue indicating underestimation, and green indicating zero deviation from the ground truth. Most regions showed minimal deviation, confirming strong local agreement with the ground truth segmentation. The maximum HD error (0.376 ± 0.085 mm for the ST and 0.409 ± 0.268 mm for the SV) corresponds to approximately a single CBCT slice thickness (0.3 mm), demonstrating that the largest discrepancies were approximately one voxel.

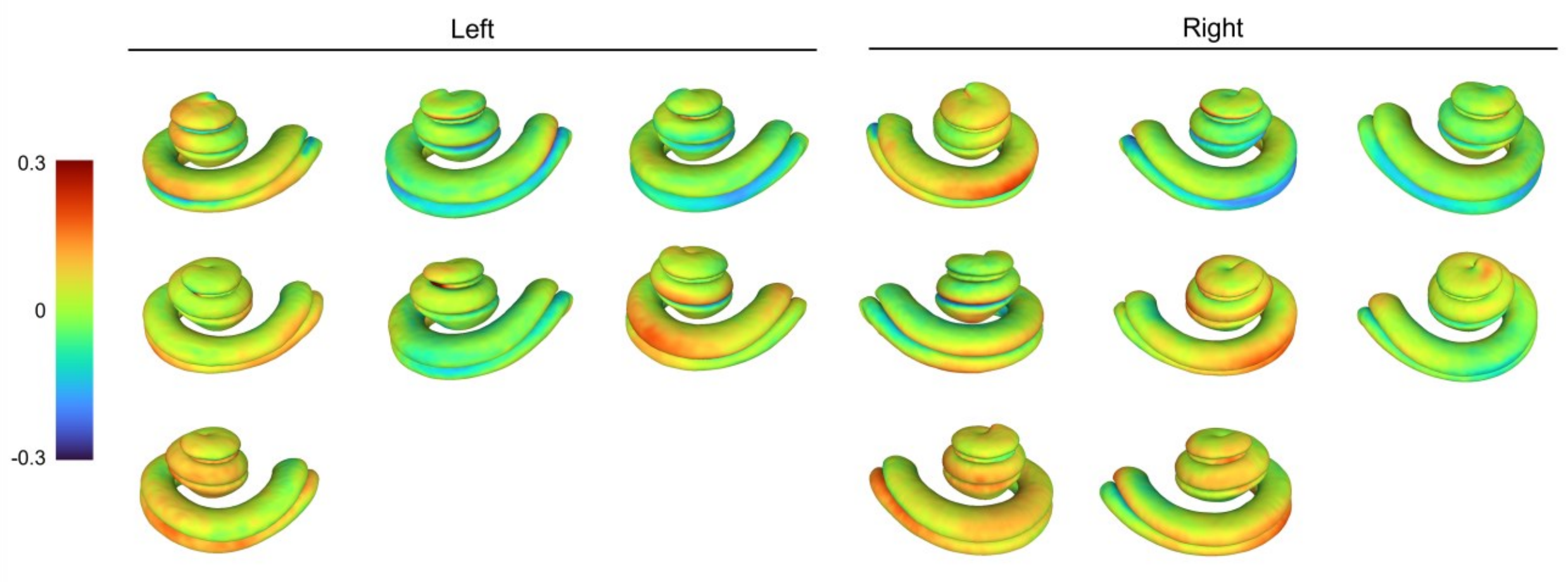

**Fig 2. Qualitative 3D representation of Euclidean distance between predicted and ground truth segmentation.** Distance for the scala tympani and scala vestibuli of each internal test sample CBCT 300 µm scan, ranging from -0.3 mm to 0.3 mm, where red represents an overestimation and blue represents an underestimation by the inference. The corresponding flipped cochlear scan segmentation is aligned with the original scan segmentation in the second column. Generated using MeshLab (v 2025.07, https://www.meshlab.net/).

Angle-specific performance was evaluated by computing two-dimensional (2D) ST DSC scores along angular depths from 45$^o$ to 810$^o$ for the CBCT 300 µm scans, as shown in Figure 3. The mean DSC remained consistently above 0.9 until 640$^o$, then decreased slightly to 0.85 at 720$^o$ and 0.79 at 810$^o$. The 2D DSC was also compared against previous techniques in the literature. Margeta *et al.*[25] was the only study to provide both 2D DSC scores and an average DSC score, reporting a ST DSC of approximately 0.75 until 270$^o$, then a decrease to approximately 0.55 at 522$^o$, beyond which no DSC values were reported. All other proposed models provided overall DSC scores, which are expected to exhibit similar trends with a high DSC in the basal turn and a decrease into the middle and apical turns due to their increased complexity. The difference between cross-sectional measurements conducted on the inference segmentations and ground truth segmentations of the test set (n=15) was analyzed using Two One-Sided Tests (TOST) for equivalence testing. For all measures, the 90% confidence interval was within the priori equivalence margin of 1 voxel (0.3 mm, 0.09 $mm^2$), except for the SV height and width from 510-550$^o$ where a maximum of -0.43 (~1.5 voxels) was observed. This suggests there is no clinically significant difference for the ST and SV metrics when comparing the inference and SR-PCI segmentations due to the consistent sub voxel equivalence (Supplemental Fig 2).

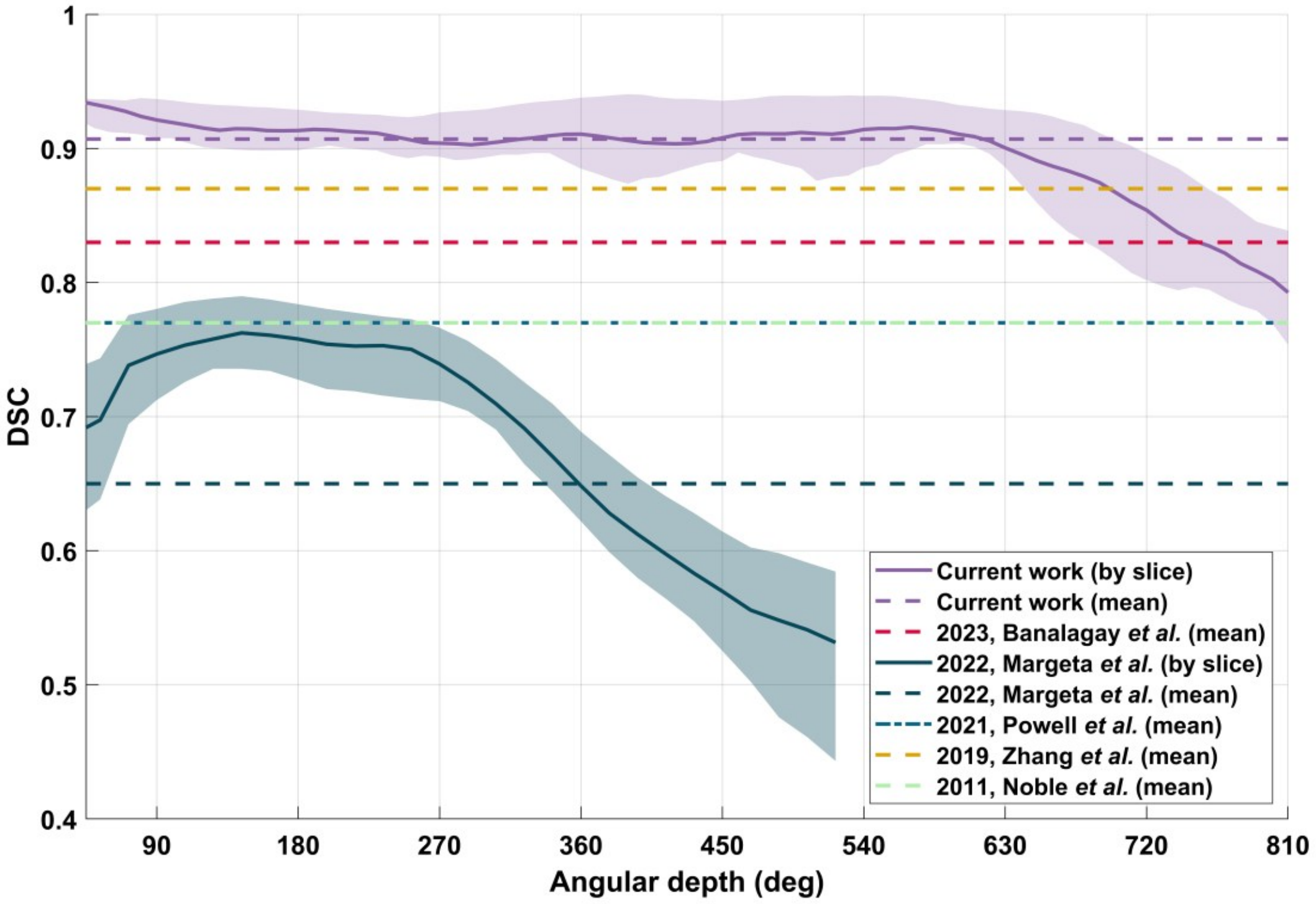

**Fig 3. Scala tympani dice similarity coefficient (DSC) as a function of angular depth.** Two-dimensional DSC scores calculated in 10$^o$ increments along the cochlea on an internal unseen test dataset of 15 CBCT scans at 300 µm. The mean DSC (purple, solid line) with the 25$^{th}$–75$^{th}$ percentile highlighted (shaded purple). For comparison, two-dimensional DSC scores are shown from the Margeta *et al.*[25] work (navy solid line; SD shaded navy) and overall DSC scores are given for all relevant studies[21–23,29] (red, blue, orange, and green dashed lines), however the falloff at high angular depths could not be graphed, as only aggregate DSC values were reported in the original studies.

*Cochlear Anatomical Metrics Evaluation*

Box plots were used to compare cochlear anatomical measurements conducted on both the ground truth and deep learning segmentation data, including the length of the ST lateral wall ($L_{ST}$), the ST volume, and the SV volume, as illustrated in Figure 4. The central line indicates the median of each measurement, and scatter points indicate individual measurements. The mean difference for the lateral wall measurements was underestimated by 0.26 ± 1.09 mm, with no statistically significant difference, suggesting unbiased and anatomically accurate predictions. The ST volume measurement exhibited an underestimation of 0.79 ± 3.20 $mm^3$, with no statistical significance shown. However, a visual underestimation of larger volumes and an overestimation of smaller volumes was observed. The SV volume measurement exhibited an overestimation of 1.71 ± 2.19 $mm^3$, with a significant difference observed (*p=0.009*). However, compared to the overall ground truth lengths and volumes, the mean error of the deep learning predictions only represented 0.71% of the mean lateral wall length, 2.49% of the mean ST volume, and 5.72% of the mean SV volume.

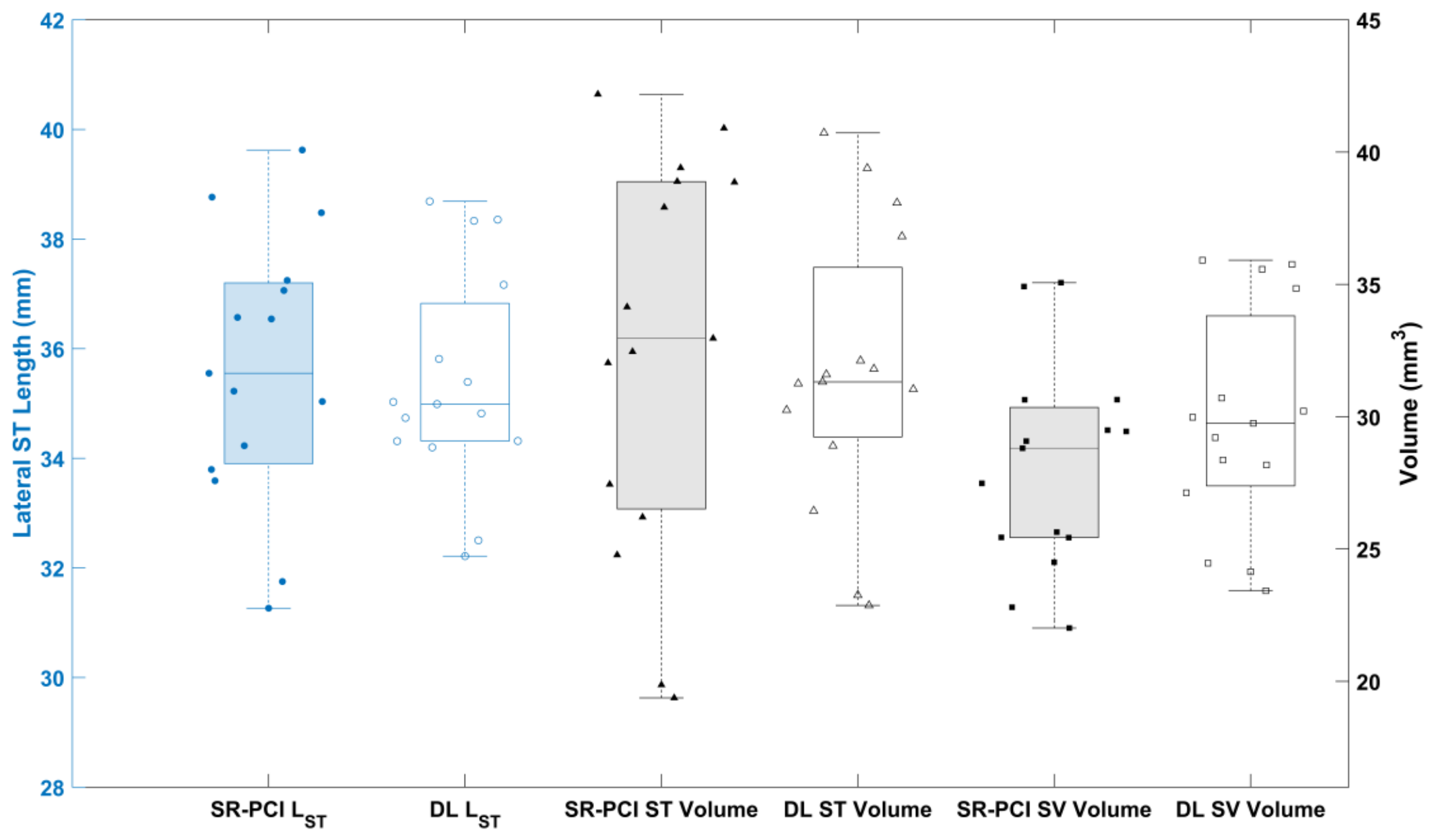

**Fig 4. Box plots comparing clinical cochlear anatomical measurements between predicted and ground truth segmentations.** Anatomical measurements included the scala tympani (ST) lateral wall length ($L_{ST}$), ST volume, and scala vestibuli (SV) volume for 15 cochleae on the CBCT 300 µm scans. Individual samples are represented for the $L_{ST}$ with circles, the ST volume with triangles, and the SV volume with squares; the ground truth synchrotron radiation phase-contrast imaging (SR-PCI) measurements are filled and the predicted deep learning (DL) measurements are unfilled. Blue boxplots correspond with the left axis in mm and black boxplots correspond with the right axis in $mm^3$. Each box displays the median as the central line, while the lower and upper box edges represent the 25th and 75th percentiles, respectively. The whiskers extend to the most extreme data points. The ground truth SR-PCI data is represented with shaded box plots, while the predicted deep learning data is represented with non-shaded box plots.

*Motion Blurring Augmented Evaluation*

The motion blurring simulated analysis of the internal testing dataset was systematically evaluated using DSC for varying levels of motion blurring as shown in Figure 5. The DSC scores remained consistently high with simulated linear blurring to 2.5 mm, demonstrating high segmentation precision. When linear blurring reached 2.5 mm, the DSC dropped by approximately 20%. Conversely, typical motion artifacts observed due to patient motion are 1.14 ± 0.84 mm on average[56]. Therefore, the segmentation model should be able to produce segmentations for all clinical scans that are of sufficient diagnostic quality, producing only marginal reductions in DSC when motion blurring is present.

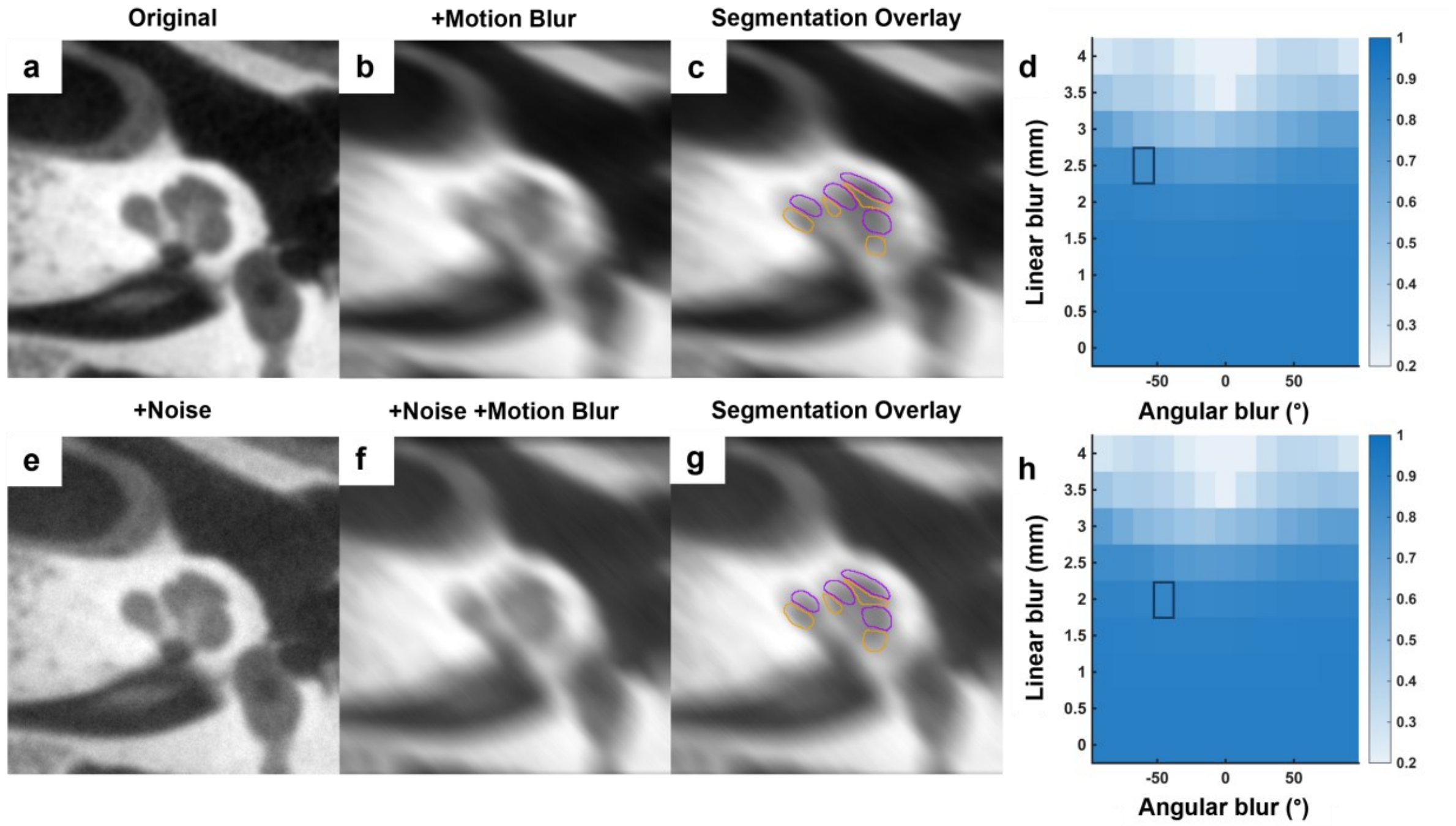

**Fig 5. Motion blur applied dice similarity coefficient (DSC) analysis of deep learning segmentation on cadaveric test dataset.** Example augmented scans are shown for **(a)** the original CBCT at 100 µm, **(b)** the original with motion blurring, **(e)** the original with noise, and **(f)** the original with both noise and motion blurring. The predicted segmentations, with the scala tympani in purple and the scala vestibuli in yellow, are overlaid in **(c, g)**. The average results comparing predicted segmentations to the ground truth using the DSC of the scala tympani in the entire dataset (n=15) are shown in **(d, h)**. The y-axis represents the linear blurring from 0 to 4 mm (0.5 mm increments), the x-axis represents angular blurring from $-90^{o}$ to $90^{o}$ ($15^{o}$ increments), and the shade of blue represents the DSC. The black box outline within the graph shows the amount of motion blurring applied in the example scans in **(d, h)**.

*Multi-institutional Deployment on Clinical Scans*

To evaluate robustness in real-world conditions, the model was tested on a multi-institutional external patient dataset comprising 501 non-pathologic clinical CT scans (55 scans from London Health Sciences Centre, Ontario, Canada; 46 scans from the University of North Carolina at Chapell Hill, NC, USA; 10 scans from Thomas Jefferson University, Philadelphia, USA; 162 scans from Ludwig-Maximilians-Universität München, Munich, Germany; and 226 scans from King Saud University, Riyadh, Saudi Arabia), as illustrated with a representative subset in Figure 6. The dataset also incorporated 22 cadaveric photon-counting CT scans (University Hospital Würzburg, Würzburg, Germany), an emerging imaging technique[57,58], further broadening the range of clinical imaging conditions represented. This ensured the model could be deployed on a wide variety of scans and modalities (Supplementary Table 1). Ground truth segmentations were unavailable for these scans, as SR-PCI is limited to cadaveric specimens. Instead, qualitative assessment, thresholding, and volume rendering were performed to examine segmentation plausibility in a clinical setting which has associated challenges such as noise, motion artefacts, and reduced contrast due to the presence of the entire skull. The model was able to produce a segmentation for all scans of diagnostic quality, and the segmentations were anatomically consistent and interpretable up to the apex in all scans. Additionally, thresholding and volume rendering on clinical resolution scans showed close bone-fluid boundary agreement between the scan and the predicted segmentation (Supplementary Fig. 3).

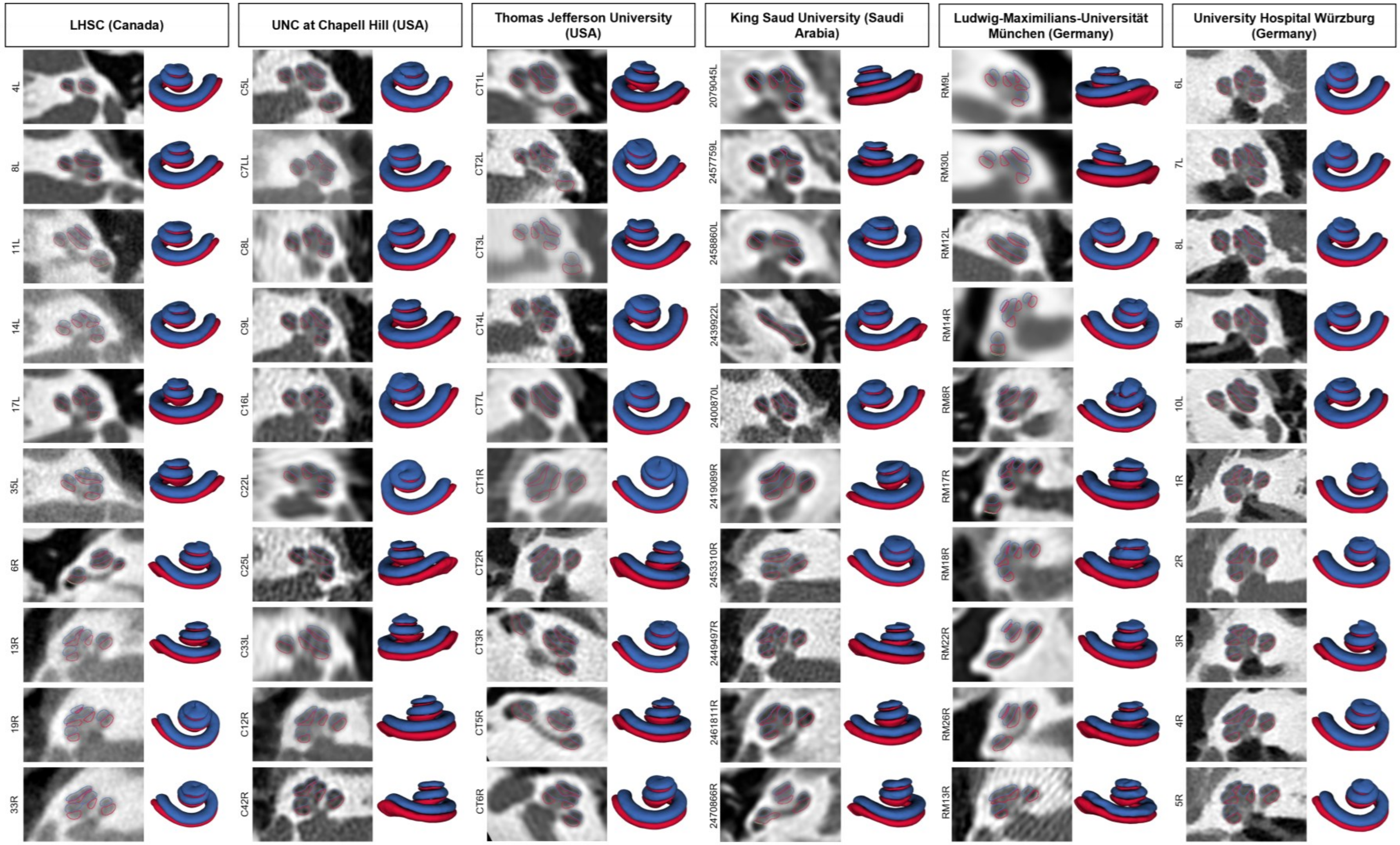

**Fig 6. Visualization of deep learning segmentations performed on the global external dataset to assess clinical plausibility.** Cross-sectional axial slices of cochlear implant recipient pre-operative scans (first five columns) and photon counting cadaveric CT scans (last column) with the scala tympani (red) and scala vestibuli (blue) segmentation outline overlayed and the 3D rendering on representative datasets from various institutions, showcasing the large variations and challenging segmentations.

## Discussion

Accurate segmentation of intracochlear structures from clinical images is essential for patient-specific cochlear implantation across preoperative, perioperative, and postoperative stages of care. Previously published segmentation models have shown success, but reduced accuracy was reported beyond the basal turn, a critical region for planning atraumatic electrode insertions and determining total scalar volume for drug delivery. Furthermore, most previous models were trained/developed using high-resolution scans obtained for research-specific purposes or scans acquired with a single clinical protocol, limiting their applicability to real-world clinical settings. The current work presents a deep learning model trained on multi-resolution clinical scans, with ground truth intracochlear segmentations derived from gold-standard synchrotron imaging, enabling robust segmentation across diverse image acquisitions. Finally, the large dataset in this study represented the large variation in cochlear size in the human population, which further increased robustness to allow for international implementation.

Compared to prior works[21–23,25,29], the proposed model achieved superior performance across all tested modalities and all objective metrics. Of the previously published works, Zhang *et al*.[29] reported the highest DSC of 0.87 and 0.86 for the ST and SV, respectively, on 300 µm CBCT scans. However, one notable limitation was the presence of cases that produced false segmentations, with anatomically implausible classifications on target scans. Noble *et al.*[22] achieved a DSC of 0.77, a mean surface distance of 0.20 mm, and maximum surface distance of 0.66 mm for the ST using paired pre-operative µCT and clinical CT scans; an improved DSC of 0.83 and surface error of 0.10 mm were reported with the increase in sample size[23]. However, µCT datasets may not have sufficient contrast, especially in delineating the apical region, and a small training dataset may restrict the model deformations therefore hindering generalizability to anatomical variations. Powell *et al.*[21] proposed a ROI-atlas intensity-based segmentation of the scala achieving a DSC of 0.77 ± 0.06, an average HD of 0.11 ± 0.01 mm, and a maximum HD of 0.78 ± 0.08 mm of the ST. However, this technique may not translate from ex-vivo to in-vivo use due to the use of aerated cochleae which impacts the intensity of the inner ear. Comparatively, the model proposed here achieved an average DSC of 0.90, an average HD of 0.008 mm, and a maximum HD 0.392 mm for the ST. This showcases the model's ability to provide segmentations of the entire cochlea on average within sub voxel accuracy, which is important for global metrics such as cochlear volume.

Margeta *et al*.[25] was the first to present 2D DSC scores in their work using deep learning for intracochlear segmentation. Their model achieved a high ST DSC of ~0.75 until 270°, then a steady decrease to a DSC of ~0.55 at 522°, past which no DSC values were presented. Furthermore, a recent subsequent study found a mean ST DSC of 0.90 when compared against histological slices at 90° and 270°[34]. Our model was able to achieve ST DSC values above 0.90 up to 640°, with a decrease to 0.79 at 810°, which is consistently higher than previously achieved across the entire cochlear length. This highlights the proposed model's ability to accurately segment the ST up to the cochlear apex, which is necessary for electrode selection for deep insertions, determining total cochlear volume for pharmacokinetics, and obtaining total cochlear duct length for frequency mapping.

Beyond classical segmentation evaluation metrics, the model achieved closely matched anatomical measurements of ST length and scalar volumes. These metrics have utility in areas such as CI electrode selection and intracochlear drug delivery via CI electrodes[50] and inner ear catheters[59]. Figure 4 compared the overall ground truth lengths and volumes, where the mean error from the deep learning predictions only represented 0.71% of the mean lateral wall length, 2.49% of the mean ST volume, and 5.72% of the mean SV volume, demonstrating accurate and unbiased prediction of the segmentation anatomical measurements. For the $L_{ST}$ measurement, a mean underestimation of 0.26 ± 1.09 mm was found, which is a sub-voxel error in clinical-resolution CT scans. The ceiling of this error is within the clinically acceptable error of 1.5 mm as proposed by Koch *et al*. when assessing cochlear length measurement errors[51,60]. The ST volume had a mean underestimation of 0.79 ± 3.20 $mm^3$ and the SV volume had a mean overestimation of 1.71 ± 2.19 $mm^3$. This may be a valuable measure for pre-operative planning since Canfarotta *et al.* [10] found that ST volume is a significant indicator of low-frequency hearing preservation within six months post implantation. The ability to accurately predict scalar volume has additional utility for planning and customizing intracochlear drug delivery with drug-eluting electrodes[61–63].

Challenges have been identified with applying automatic segmentation methods to clinical scans, primarily related to poor resolution and imaging conditions[25]. Patient scans can exhibit noise variations across institutions due to differences in scanner type, acquisition settings, and reconstruction settings. Additionally, clinical imaging of the cochlea has inherent challenges due to the size of the structures relative to the imaging artefacts, as well as the reduced contrast from the presence of the entire temporal bone, creating barriers to translating empirical methods developed using cadaveric data into clinical practice. To address these challenges, this work incorporated cadaveric data from multiple scanners, protocols, and resolutions, along with clinically driven augmentation strategies such as blurring and noise, to improve model robustness against scanner variability and patient-specific artifacts. Furthermore, cadaveric clinical images with and without fluid were acquired and random intensity shifting was applied prior to training to reduce reliance on intensity-based features alone. In addition, motion-artefacts were simulated on cadaveric images to evaluate the model's robustness against real-world challenges. The model was able to accommodate linear motion up to 2.5 mm, which is quite robust given the average cochlear base length (A-value) is only 9 mm[47], where moderate shifts can distort anatomical boundaries and compromise segmentation performance. When linear motion blurring exceeded 2.5 mm, the DSC dropped by approximately 20%. However, typical motion artifacts due to patient motion are 1.14 ± 0.84 mm on average[56]. Therefore, the segmentation model should capture the majority of clinical scans performed that maintain diagnostic integrity, producing only marginal reductions in DSC when blurring is present.

Previous studies have been limited by the use of scans from a single center, obtained with the same scanner and protocol, for both the validation and training datasets. In addition, the small datasets may not have represented the large cochlear variation present in humans. This reduces the real-world applicability of these models to the global population, where variability exists in scanner type, imaging protocols, patient anatomy, and positioning/movement. Therefore, this study aimed to validate the automatic segmentations across scans obtained from various institutions spanning four countries. Since ground truth SR-PCI data cannot be obtained *in vivo*, model performance on clinical data was evaluated through qualitative assessment, thresholding, and volume rendering. All resulting segmentations were anatomically viable, demonstrating the model's ability to generalize across diverse clinical conditions. Furthermore, there are already initial automated segmentation tools available in the current clinical workflows[64], which demonstrates the clinical feasibility of adopting sophisticated deep learning approaches into established pipelines for precise and individualized patient care.

Intracochlear trauma has been associated with delayed residual hearing loss and inferior post-operative hearing outcomes [65–68]. Figure 1 demonstrates that the model accurately segments both scala, while also excluding the modiolus, spiral ligament, and basilar membrane. Therefore, by determining electrode position using fused pre- and post-operative scans, future work may focus on accurately assessing intracochlear trauma in patient scans due to modiolar wall injury, lateral wall injury, or scalar translocation.

A limitation of the model is that it is unable to segment abnormal cochleae, such as malformations (e.g. hypoplasia or common cavities) or advanced cochlear otosclerosis, as these pathologies were not included in the training dataset. However, these could be incorporated into future work if suitable imaging databases become available. Additionally, magnetic resonance imaging (MRI) could be incorporated in the future to garner additional intracochlear information and enhance the model's utility.

In this work, over 100 unique cochleae with paired SR-PCI segmentations and clinical-resolution scans were used to train an automatic scalar segmentation model for various CT manufacturers, scan types, and resolutions. Objective metrics showed high accuracy through to the apex of the cochlea, which has been previously unattainable. Additionally, strong agreement was found between the predicted and ground truth segmentations across various anatomical measurements.

## Methods

### *Image Acquisition and Pre-processing*

One hundred and eight human cadaveric cochleae (55 left, 53 right; 59 female, 41 male, 8 with unknown sex data) were obtained as part of the body bequeathal program at Western University, London, Ontario, Canada in accordance with the Anatomy Act of Ontario and Western's Committee for Cadaveric Use in Research (approval #122611). Specimens were chosen to represent the wide range of cochlear morphologies that are seen in the population, characterized by scala tympani (ST) as previously reported[7].

Synchrotron radiation phase-contrast images (SR-PCI) CT scans were obtained of all 108 cadaveric cochleae. SR-PCI scanning was performed at the Canadian Light Source Inc. Biomedical Imaging and Therapy beamline in Saskatchewan, Canada. The resulting SR-PCI CT scans had an isotropic voxel spacing of 8-9 µm; the detailed imaging procedure has been previously described[48]. Up to four unique clinical CT scanners were used to image each cochlea, with various protocols and resolutions used for reconstruction. These various clinical scanners and reconstruction techniques resulted in up to nine unique CT volumes for each cochlea. The scanners, reconstruction techniques, and sample sizes are summarized with example CT slices in Table 2.

All images were processed using 3D Slicer (version 5.6.1, https://www.slicer.org/), an open-source medical imaging software[69]. All scans were aligned and registered to each other in a clinical view using a three-step registration process: (1) atlas-based landmark alignment of the SR-PCI images to a "clinical view" (representative of a supine position); (2) manual rigid landmark registration of clinical scans to the SR-PCI reference in the clinical view; and (3) intensity-based rigid registration with the Elastix and Brains modules in 3D Slicer[70,71]. All scans and label maps were resampled to a uniform voxel spacing. To reduce data requirements, a fiducial was manually placed on the cochlea in each scan and a bounding box centered on this landmark was automatically applied for cropping, ensuring the entire cochlea was enclosed.

SR-PCI data was used to establish ground truth segmentations of intracochlear anatomy. The multi-class ground truth segmentation label maps were generated to include the ST and scala vestibuli (SV) (including the scala media) using semi-automatic segmentation followed by manual correction on SR-PCI data. Consensus ground truth segmentations were achieved by three observers for all cases.

Data augmentation was performed on all clinical-resolution reconstructions to increase the training dataset size and enhance robustness to different imaging characteristics. Augmentation techniques included mirroring, additive Gaussian noise at two intensity levels, and Gaussian blurring[28].

**Table 2. The number of cochleae scanned with each imaging modality and protocol.** Scan type, resolution, and acquisition details are provided, along with an example mid-modiolar slice of the same specimen for each scan type. A single value for the resolution indicates isotropy and three values indicate anisotropy.

| # of Samples Scanned | Modality/ Scan Type | Resolution/ Protocol | Mid-modiolar slice |
|---|---|---|---|
| 108 | SR-PCI | 0.008 - 0.009 mm | 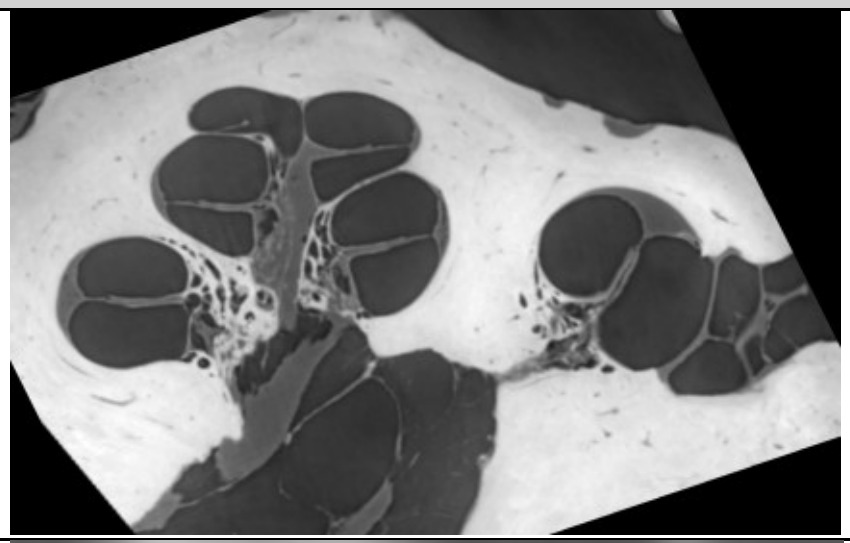 |
| 108 | Xoran MiniCAT CBCT | 0.1 mm | 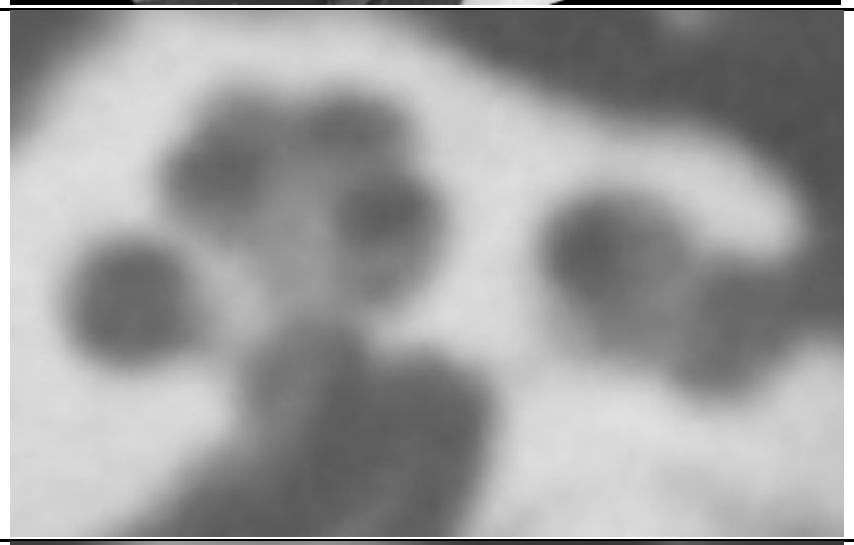 |
| 108 | Xoran MiniCAT CBCT | 0.3 mm | 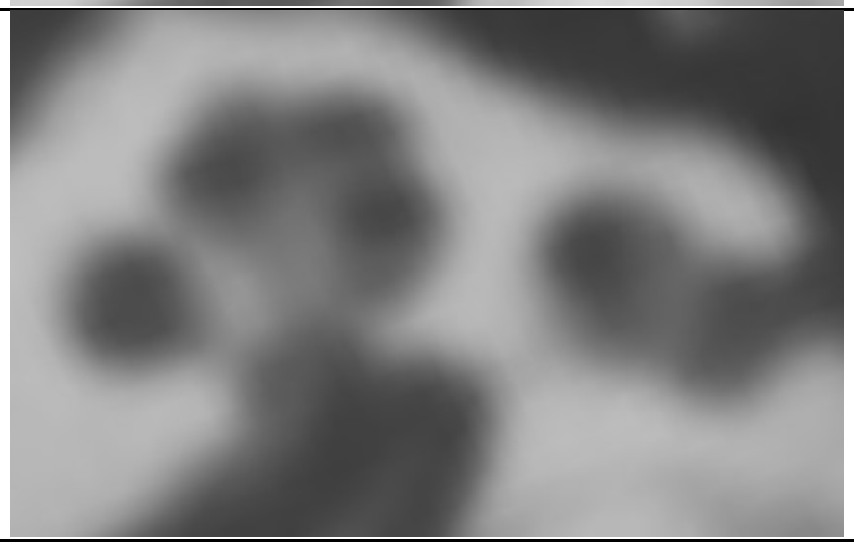 |
| 42 | GE Discovery CT750 HD Helical CT | 0.488 x 0.488 x 0.625 mm, Temporal Bone Protocol | 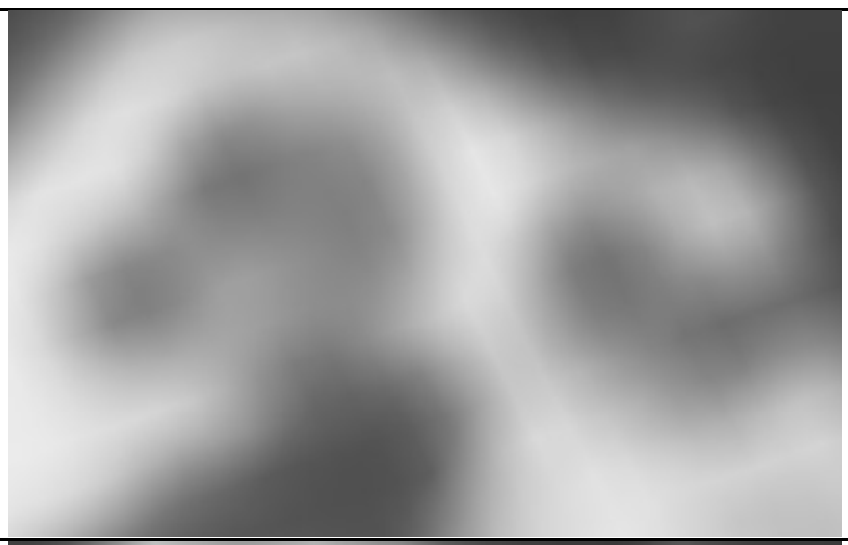 |
| 40 | GE Discovery CT750 HD Helical CT | 0.488 x 0.488 x 0.625 mm, Soft Tissue Protocol | 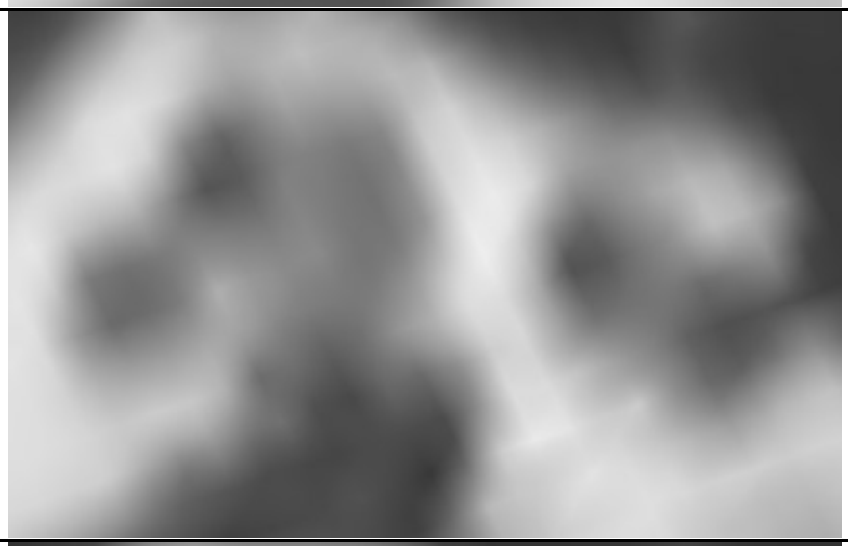 |
| 103 | Canon Aquilion ONE Helical CT | 0.591 x 0.591 x 0.5 – 0.789 x 0.789 x 0.5 mm, Temporal Bone Protocol | 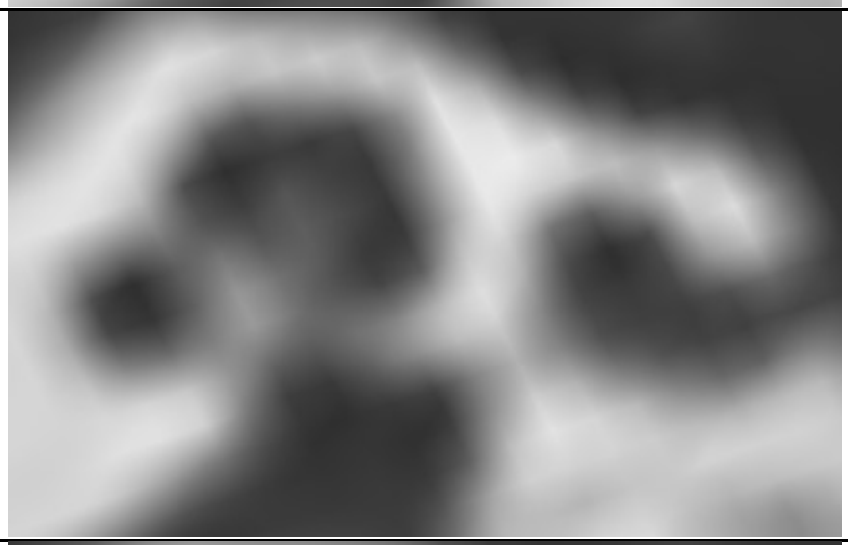 |
| 103 | Canon Aquilion ONE Helical CT | 0.591 x 0.591 x 0.5 – 0.789 x 0.789 x 0.5 mm, Soft Tissue Protocol | 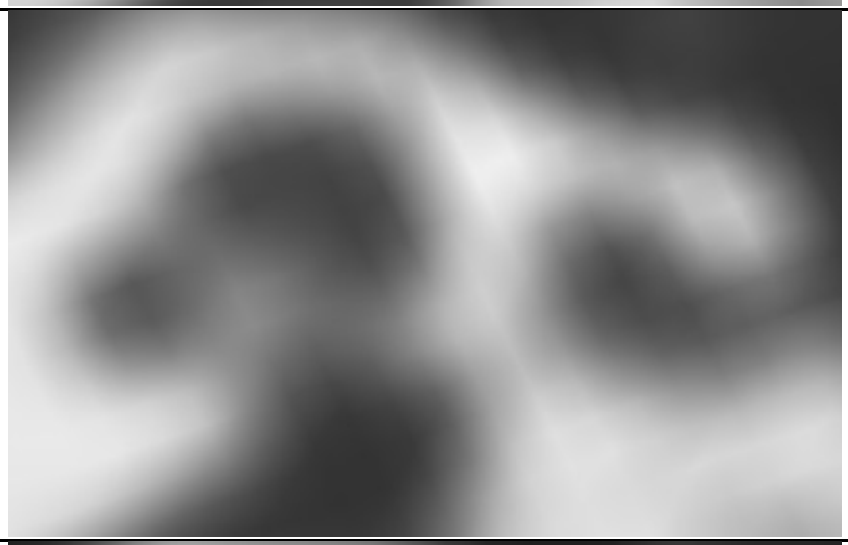 |
| 100 | Siemens AXIOM-Artis Angiography CT | 0.48 mm, Normal Protocol | 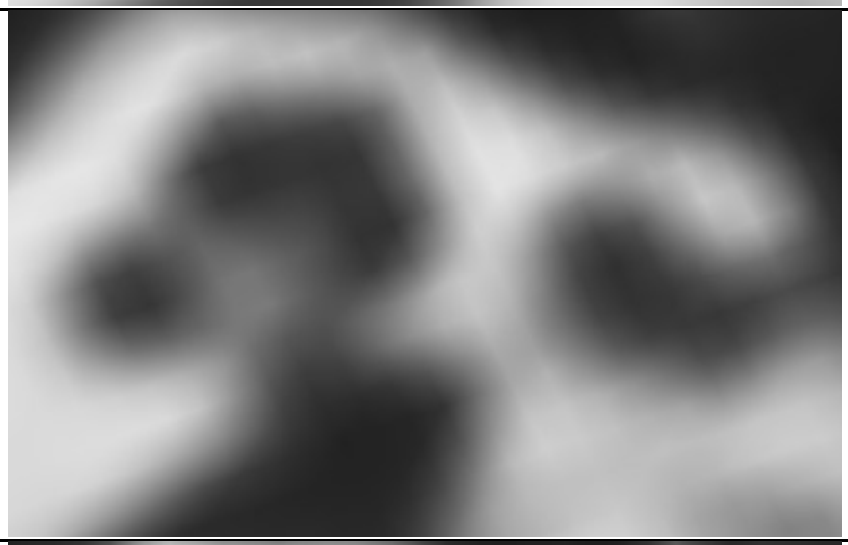 |
| 100 | Siemens AXIOM-Artis Angiography CT | 0.17 – 0.45 mm, Sharp Protocol | 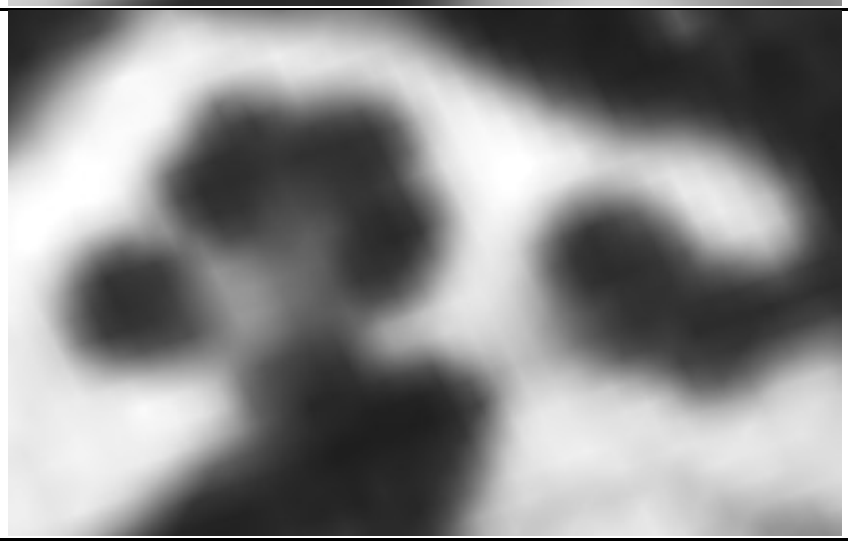 |

*Global Objective Metrics Evaluation*

Model performance was evaluated on the test set using global and local metrics to compare the deep learning predicted segmentations and ground truth SR-PCI segmentations of the same cochlea. Global metrics included Dice similarity coefficient (DSC) and maximum and average Hausdorff distance (HD) for the ST and SV. DSC quantifies spatial overlap between predicted and reference segmentations, and HD captures boundary errors.

*Local Objective Metrics Evaluation*

Local performance was assessed using Euclidean distance maps between predicted and ground truth segmentations to determine the deviation at each location on the surface of the predicted segmentation. Additionally, local performance was assessed using 2D reformatted mid-modiolar slices to obtain cross-sectional representations of the scalae at progressive angular depths. Two-dimensional DSC values were computed at $10^{o}$ increments along the ST and by comparing the cross-sections of the predicted scala and reference SR-PCI scala. Additional cross-sectional measurement differences, including the diameter, area, width, and height of the ST and SV[7], were analyzed using equivalence testing and are provided in the supplementary materials. Equivalence was assessed using Two One-Sided Tests (TOST)[72], evaluating whether the 90% confidence interval of the mean difference at each angle for each metric was within a priori equivalence margin of 1 voxel (0.3 mm, 0.09 $mm^2$). Sub voxel equivalence would suggest clinically acceptable equivalent measurements between the deep learning and SR-PCI segmentations. These metrics help elucidate segmentation reliability at progressive locations from the cochlear base to apex.

*Cochlear Anatomical Metrics Evaluation*

The total ST and SV volumes, and the total length of the scala tympani lateral wall ($L_{ST}$) were compared between the predicted segmentations and reference SR-PCI segmentations. Volumetric measurements were obtained by voxel summation in each class and converted to $mm^3$. The $L_{ST}$ was defined as the ST length from the center of the RW to the apex along the lateral wall. This was measured in cochlear view using a standard coordinate system to conduct anatomical cochlear measurements[73]. Agreement between the predicted and reference measurements was evaluated using box plots for the 300 µm CBCT scans. The statistical difference between the ground truth and predicted value for each measurement was evaluated using a t-test. These plots can illustrate sources of systematic bias and identify over- or underestimation trends.

*Motion Blurring Augmented Evaluation*

To assess the model’s robustness under conditions resembling patient imaging artefacts associated with anisotropy, augmentations simulating motion blurring were applied to the cadaveric scans. This augmentation was not applied during training and was used exclusively for evaluation. Motion blurring artefacts were generated using a distance-weighted linear averaging along a line segment with a length from 0 to 4 mm (in 0.5 mm increments) and angles from $-90^{o}$ to $90^{o}$ counterclockwise (in $15^{o}$ increments), in-plane to the original and noise augmented images. The motion-augmented images were processed by the trained ensemble model, and segmentation performance was quantified using the DSC. This analysis provides a systematic evaluation of the model's robustness to clinically plausible image distortions and its applicability in real-world scenarios where image-related artefacts are possible.

*Multi-institutional Deployment on Clinical Data*

To evaluate the model's ability to generalize in real-world conditions, the deep learning model was tested on a temporal bone imaging database (Research Ethics Board approval #112296, #120513, and #120490). This database was comprised of 501 anonymized pre-operative clinical scans acquired for cochlear implant candidates and 22 anonymized cadaveric clinical scans across six international sites (55 scans from London Health Sciences Centre, Ontario, Canada; 46 scans from the University of North Carolina at Chapell Hill, NC, USA; 10 scans from Thomas Jefferson University, Philadelphia, USA; 162 scans from Ludwig-Maximilians-Universität München, Munich, Germany; 22 scans from the University Hospital Würzburg, Würzburg, Germany; and 226 scans from King Saud University, Riyadh, Saudi Arabia). Ground truth segmentations were unavailable for these scans, as synchrotron imaging is limited to cadaveric specimens. Therefore, qualitative assessment, thresholding, and volume rendering were performed to assess segmentation plausibility against clinical scans with relevant challenges such as noise, motion artefacts, and anatomical variability.

<u>Data availability</u>

Human imaging data and derivative algorithms are not publicly available due to ethics restrictions and data privacy laws. However, access may be available for academic purposes via ethics approval and data sharing agreements by sending a request to the corresponding author.

<u>Code availability</u>

The baseline deep learning code used in this study is implemented in the Auto3DSeg framework within MONAI, available at https://github.com/Project-MONAI.

<u>Acknowledgements</u>

Part of the research described in this paper was performed at the Canadian Light Source, a national research facility of the University of Saskatchewan, which is supported by the Canada Foundation for Innovation (CFI), the Natural Sciences and Engineering Research Council (NSERC), the National Research Council (NRC), the Canadian Institutes of Health Research (CIHR), the Government of Saskatchewan, and the University of Saskatchewan.

Lauren Siegel provided manuscript editing and review.

<u>Author contributions</u>

A.M. wrote the main manuscript text, planned the study, and analyzed the results; H.L. and S.K.A. conceptualized the research methodology, reviewed the article, optimized and applied SR-PCI to imaging of the cochlea, and performed relevant data processing; D.N., N.S., S.P., A.A., K.D.B, A.H., J.B.H, J.M., and K.R. aided in acquiring/providing clinical scans for validation.

## Disclosures

K.D.B, A.H., J.M., K.R., and S.K.A. are on the Surgical Advisory Board of MED-EL GmBH. A.M. is a PhD student in the Auditory Biophysics Laboratory at Western University and completed an internship at MED-EL GmBH.

MED-EL GmBH has previously provided a donation to Western University and supported a MITACS grant. However, the study was completed independently; industry had no role in the study design; data collection, analysis, or interpretation; manuscript preparation; or the decision to submit for publication.

All remaining authors have no conflict of interest.

## Additional information

**Correspondence** and requests for materials should be addressed to H.L. or S.K.A.

References:

1. World Health Organization. Deafness and hearing loss. *https://www.who.int/health-topics/hearing-loss* (2021).

2. Davis, A. C. & Hoffman, H. J. Hearing loss: Rising prevalence and impact. *Bull. World Health Organ.* **97**, 646 (2019).

3. Roland, J. T. A model for cochlear implant electrode insertion and force evaluation: Results with a new electrode design and insertion technique. *Laryngoscope* **115**, 1325–1339 (2005).

4. Schurzig, D. *et al.* Virtual Cochlear Implantation for Personalized Rehabilitation of Profound Hearing Loss. *Hear. Res.* **492**, 108687 (2022).

5. Avci, E., Nauwelaers, T., Lenarz, T., Hamacher, V. & Kral, A. Variations in microanatomy of the human cochlea. *Journal of Comparative Neurology* **522**, 3245–3261 (2014).

6. Erixon, E., Högstorp, H., Wadin, K. & Rask-Andersen, H. Variational anatomy of the human cochlea: Implications for cochlear implantation. *Otology and Neurotology* **30**, 14–22 (2009).

7. Micuda, A., Li, H., Rask-Andersen, H., Ladak, H. M. & Agrawal, S. K. Morphologic Analysis of the Scala Tympani Using Synchrotron: Implications for Cochlear Implantation. *Laryngoscope* **134**, 2889–2897 (2024).

8. Rask-Andersen, H. *et al.* Human cochlea: Anatomical characteristics and their relevance for cochlear implantation. *Anatomical Record* **295**, 1791–1811 (2012).

9. Alzhrani, F. *et al.* Preoperative Imaging for Cochlear Implantation: A Global Consensus. *Otolaryngology–Head and Neck Surgery* https://doi.org/10.1002/ohn.1361 (2026) doi:10.1002/ohn.1361.

10. Canfarotta, M. W., Dillon, M. T., Selleck, A. M. & Brown, K. D. Scala Tympani Volume Influences Initial 6-Month Hearing Preservation With Lateral Wall Electrode Arrays. *Laryngoscope* **135**, 1781–1787 (2025).

11. Bircher, B. *et al.* A clinical shift toward personalized cochlear implantation: Using preoperative planning to optimize insertion depth. *American Journal of Otolaryngology - Head and Neck Medicine and Surgery* **47**, (2025).

12. Räth, M., Schurzig, D., Timm, M. E., Lenarz, T. & Warnecke, A. Correlation of Scalar Cochlear Volume and Hearing Preservation in Cochlear Implant Recipients with Residual Hearing. *Otol. Neurotol.* **45**, 256–265 (2024).

13. Müller-Graff, F.-T. *et al.* Position Control of Flexible Electrodes With Regard to Intracochlear Structure Preservation and Hearing Outcomes: A Retrospective Study With Implementation of the Electrode Contact View. *Otology & Neurotology* **46**, e307–e315 (2025).

14. O'Connell, B. P. *et al.* Electrode Location and Angular Insertion Depth Are Predictors of Audiologic Outcomes in Cochlear Implantation. *Otology and Neurotology* **37**, 1016–1023 (2016).

15. Canfarotta, M. W. *et al.* Effects of Insertion Depth and Modiolar Proximity on Cochlear Implant Speech Recognition Outcomes With a Precurved Electrode Array. *Otology & Neurotology* **46**, 272–278 (2025).

16. Dillon, M. T. *et al.* Influence of the Frequency-to-Place Function on Recognition with Place-Based Cochlear Implant Maps. *Laryngoscope* **133**, 3540–3547 (2023).

17. Kurz, A., Herrmann, D., Hagen, R. & Rak, K. Using Anatomy-Based Fitting to Reduce Frequency-to-Place Mismatch in Experienced Bilateral Cochlear Implant Users: A Promising Concept. *J. Pers. Med.* **13**, 1109 (2023).

18. Kurz, A. *et al.* Anatomy-based fitting improves speech perception in noise for cochlear implant recipients with single-sided deafness. *European Archives of Oto-Rhino-Laryngology* **282**, 467–479 (2025).

19. Siebrecht, M., Briaire, J. J., Verbist, B. M., Kalkman, R. K. & Frijns, J. H. M. Automated segmentation of clinical CT scans of the cochlea and analysis of the cochlea's vertical profile. *Heliyon* **10**, (2024).

20. Andersen, S. A. W. *et al.* Segmentation of Temporal Bone Anatomy for Patient-Specific Virtual Reality Simulation. *Annals of Otology, Rhinology and Laryngology* **130**, 724–730 (2021).

21. Powell, K. A. *et al.* Atlas-based segmentation of cochlear microstructures in cone beam CT. *Int. J. Comput. Assist. Radiol. Surg.* **16**, 363–373 (2021).

22. Noble, J. H., Labadie, R. F., Majdani, O. & Dawant, B. M. Automatic segmentation of intracochlear anatomy in conventional CT. *IEEE Trans. Biomed. Eng.* **58**, 2625–2632 (2011).

23. Banalagay, R. A., Labadie, R. F. & Noble, J. H. Validation of active shape model techniques for intracochlear anatomy segmentation in computed tomography images. *Journal of Medical Imaging* **10**, 044003–044003 (2023).

24. Neves, C. A., Chemaly, T. El., Fu, F. & Blevins, N. H. Deep Learning Method for Rapid Simultaneous Multistructure Temporal Bone Segmentation. *Otolaryngology–Head and Neck Surgery* **170**, 1570–1580 (2024).

25. Margeta, J. *et al.* A Web-Based Automated Image Processing Research Platform for Cochlear Implantation-Related Studies. *J. Clin. Med.* **11**, 6640 (2022).

26. Hussain, R. *et al.* Anatomical Variations of the Human Cochlea Using an Image Analysis Tool. *J. Clin. Med.* **12**, 509 (2023).

27. Heutink, F. *et al.* Multi-Scale deep learning framework for cochlea localization, segmentation and analysis on clinical ultra-high-resolution CT images. *Comput. Methods Programs Biomed.* **191**, 105387 (2020).

28. Nikan, S. *et al.* PWD-3DNet: a deep learning-based fully-automated segmentation of multiple structures on temporal bone CT scans. *IEEE Transactions on Image Processing* **30**, 739–753 (2021).

29. Zhang, D. *et al.* Two-level training of a 3D U-Net for accurate segmentation of the intra-cochlear anatomy in head CTs with limited ground truth training data. in *Medical Imaging 2019: Image Processing* vol. 10949 45–52 (SPIE, 2019).

30. Yoo, T. W., Yeo, C. D., Lee, E. J. & Oh, I. S. Cross-channel feature transfer 3D U-Net for automatic segmentation of the perilymph and endolymph fluid spaces in hydrops MRI. *Comput. Biol. Med.* **197**, (2025).

31. Noble, J. H. & Dawant, B. M. Automatic graph-based localization of cochlear implant electrodes in CT. *Med Image Comput Comput Assist Interv* **9350**, 152–159 (2015).

32. Zhao, Y., Dawant, B. M., Labadie, R. F. & Noble, J. H. Automatic Localization of Cochlear Implant Electrodes in CT. *Med Image Comput Comput Assist Interv* **17**, 331–338 (2014).

33. Tawfik, K. O. *et al.* Cochlear Implantation of Slim Precurved Arrays Using Automatic Preoperative Insertion Plans. *Otology and Neurotology* **46**, 862–870 (2025).

34. Hussain, R., Patou, F., Calixto, R., Matti, I.-M. & Dietz, A. A histological validation of an automatic cochlear implant electrode placement assessment from clinical computed tomography images. *Cochlear Implants Int.* **26**, 341–348 (2025).

35. Schreier, A. *et al.* Individualized cochlear implantation – first experience with a new 34 mm electrode for patients with very long cochleae. *Laryngo-Rhino-Otologie* **103**, (2024).

36. O'Rourke, S. P. *et al.* Association of Tonotopic Mismatch and Cochlear Implant Use on Outcomes for Adults With Unilateral Hearing Loss. *Laryngoscope* (2025).

37. Nassiri, A. M. *et al.* Hearing Preservation Outcomes Using a Precurved Electrode Array Inserted with an External Sheath. *Otology and Neurotology* **41**, 33–38 (2020).

38. Canfarotta, M. W. *et al.* Frequency-To-Place Mismatch: Characterizing Variability and the Influence on Speech Perception Outcomes in Cochlear Implant Recipients. *Ear Hear.* **41**, 1349–1361 (2020).

39. Canfarotta, M. W. *et al.* Long-Term Influence of Electrode Array Length on Speech Recognition in Cochlear Implant Users. *Laryngoscope* **131**, 892–897 (2021).

40. Weller, T., Timm, M. E., Lenarz, T. & Büchner, A. Cochlear coverage with lateral wall cochlear implant electrode arrays affects post-operative speech recognition. *PLoS One* **18**, e0287450 (2023).

41. Fan, T. *et al.* Effect of Electrode Insertion Angle on Cochlear Implantation Outcomes in Adult and Children Patients with Sensorineural Hearing Loss. *Oxid. Med. Cell. Longev.* **2022**, 9914716 (2022).

42. O'Connell, B. P. *et al.* Insertion depth impacts speech perception and hearing preservation for lateral wall electrodes. *Laryngoscope* **127**, 2352–2357 (2017).

43. Buchman, C. A. *et al.* Influence of Cochlear Implant Insertion Depth on Performance: A Prospective Randomized Trial. *Otology & Neurotology* **35**, 1773–1779 (2014).

44. Breitsprecher, T. M. *et al.* Effect of Cochlear Implant Electrode Insertion Depth on Speech Perception Outcomes: A Systematic Review. *Otology & Neurotology Open* **3**, e045 (2023).

45. Triche, B. L. *et al.* Recognizing and minimizing artifacts at ct, mri, us, and molecular imaging. *Radiographics* **39**, 1017–1018 (2019).

46. Hardy, M. The length of the organ of Corti in man. *American Journal of Anatomy* **62**, 291–311 (1938).

47. Human-Baron, R. & Hanekom, T. Unpacking the terminology used in human cochlear dimension methodologies. *Translational Research in Anatomy* **35**, (2024).

48. Elfarnawany, M. *et al.* Micro-CT versus synchrotron radiation phase contrast imaging of human cochlea. *J. Microsc.* **265**, 349–357 (2017).

49. Li, H. *et al.* Synchrotron Radiation-Based Reconstruction of the Human Spiral Ganglion: Implications for Cochlear Implantation. *Ear Hear.* **41**, 173–181 (2018).

50. Helpard, L. *et al.* An Approach for Individualized Cochlear Frequency Mapping Determined from 3D Synchrotron Radiation Phase-Contrast Imaging. *IEEE Trans. Biomed. Eng.* **68**, 3602–3611 (2021).

51. Koch, R. W., Elfarnawany, M., Zhu, N., Ladak, H. M. & Agrawal, S. K. Evaluation of Cochlear Duct Length Computations Using Synchrotron Radiation Phase-Contrast Imaging. *Otology and Neurotology* **38**, e92–e99 (2017).

52. Mei, X. *et al.* Vascular Supply of the Human Spiral Ganglion: Novel Three-Dimensional Analysis Using Synchrotron Phase-Contrast Imaging and Histology. *Sci. Rep.* **10**, (2020).

53. Li, H. *et al.* Three-dimensional tonotopic mapping of the human cochlea based on synchrotron radiation phase-contrast imaging. *Sci. Rep.* **11**, (2021).

54. Chen, J. M. *et al.* Synchrotron-Based Trauma Assessment of Robotic Electrode Insertions in Cochlear Implantation. *Laryngoscope* **135**, 3867–3876 (2025).

55. Cardoso, M. J. *et al.* MONAI: An open-source framework for deep learning in healthcare. *arXiv preprint* http://arxiv.org/abs/2211.02701 (2022).

56. Hanzelka, T. *et al.* Movement of the patient and the cone beam computed tomography scanner: Objectives and possible solutions. *Oral Surg. Oral Med. Oral Pathol. Oral Radiol.* **116**, 769–773 (2013).

57. Rak, K. *et al.* The Photon-Counting CT Enters the Field of Cochlear Implantation: Comparison to Angiography DynaCT and Conventional Multislice CT. *Otology and Neurotology* **45**, 662–670 (2024).

58. Batts, S., Pham, N., Tearney, G. & Stankovic, K. M. The State of High-Resolution Imaging of the Human Inner Ear: A Look Into the Black Box. *Advanced Science* **12**, (2025).

59. Skarżyńska, M. B. *et al.* Local delivery of steroids to inner ear via medical device INCAT (the Inner Ear Catheter) in partial deafness patients during cochlear implantation–preliminary results and a feasibility study. *Expert Opin. Drug Deliv.* **22**, 599–607 (2025).

60. Helpard, L. W., Rohani, S. A., Ladak, H. M. & Agrawal, S. K. Evaluation of Cochlear Duct Length Measurements from a 3D Analytical Cochlear Model Using Synchrotron Radiation Phase-Contrast Imaging. *Otology and Neurotology* **41**, e21–e27 (2020).

61. Maimaitikelimu, X. *et al.* Rational Design of Inner Ear Drug Delivery Systems. *Advanced Science* **12**, 2410568 (2025).

62. Prenzler, N. *et al.* Cochlear implantation with a dexamethasone-eluting electrode array: First-in-human safety and performance results. *Hear. Res.* **461**, 109255–109255 (2025).

63. Dong, X., Li, H. & Zuo, W. Local drug delivery systems for the inner ear. *RPS Pharmacy and Pharmacology Reports* **2**, (2023).

64. OTOPLAN Software. https://www.medel.pro/products/otoplan.

65. Quesnel, A. M. *et al.* Delayed loss of hearing after hearing preservation cochlear implantation: Human temporal bone pathology and implications for etiology. *Hear. Res.* **333**, 225–234 (2016).

66. Wanna, G. B. *et al.* Impact of electrode design and surgical approach on scalar location and cochlear implant outcomes. *Laryngoscope* **124**, S1–S7 (2014).

67. Holden, L. K. *et al.* Factors Affecting Open-Set Word Recognition in Adults With Cochlear Implants. *Ear Hear.* **34**, 342–360 (2013).

68. Ishiyama, A. *et al.* Post hybrid cochlear implant hearing loss and endolymphatic hydrops. *Otology and Neurotology* **37**, 1516–1521 (2016).

69. Fedorov, A. *et al.* 3D Slicer as an image computing platform for the Quantitative Imaging Network. *Magn. Reson. Imaging* **30**, 1323–1341 (2012).

70. Klein, S., Staring, M., Murphy, K., Viergever, M. A. & Pluim, J. P. W. Elastix: A Toolbox for Intensity-Based Medical Image Registration. *IEEE Trans. Med. Imaging* **29**, 196–205 (2010).

71. Johnson, H., Harris, G. & Williams, K. BRAINSFit: Mutual Information Registrations of Whole-Brain 3D Images, Using the Insight Toolkit. *Insight J.* **57**, 1–10 (2007).

72. Anisha. TOST(sample1, sample2, d1, d2, alpha). *MATLAB Central File Exchange* https://www.mathworks.com/matlabcentral/fileexchange/63204-tost-sample1-sample2-d1-d2-alpha (2026).

73. Verbist, B. M. *et al.* Consensus panel on a cochlear coordinate system applicable in histologic, physiologic, and radiologic studies of the human cochlea. *Otology and Neurotology* **31**, 722–730 (2010).

Supplementary

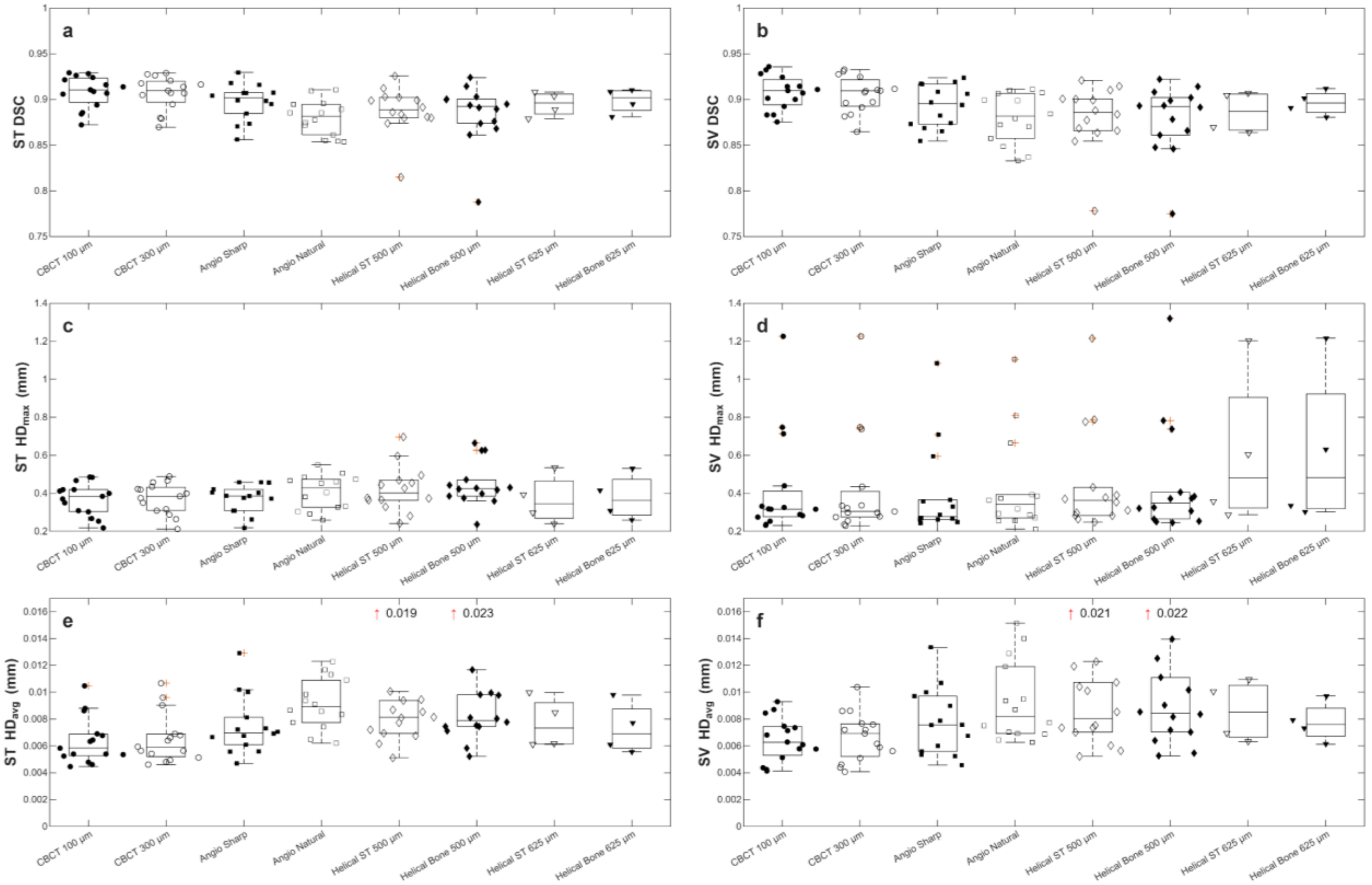

**Supplementary Fig 1. Objective measures comparing model inferences to ground truth segmentations for the scala tympani (ST) and scala vestibuli (SV).** Metrics include overall Dice similarity coefficient (DSC) **(a, b)**, maximum Hausdorff distance ($HD_{max}$) **(c, d)**, and average Hausdorff distance ($HD_{avg}$) **(e, f)** for the ST and SV respectively. Box plots are shown for each scan type with scatter points overlayed to represent individual samples. Marker style represents the same acquisition (circle, square, diamond, triangle), and the fill represents different protocols. The median is indicated by the central line in each boxplot, with the lower and upper bounds representing the 25th and 75th percentiles. Whiskers span the most extreme non-outlier observations, and outliers are plotted individually as red '+' markers or with '↑' markers if they extend beyond the axis limit.

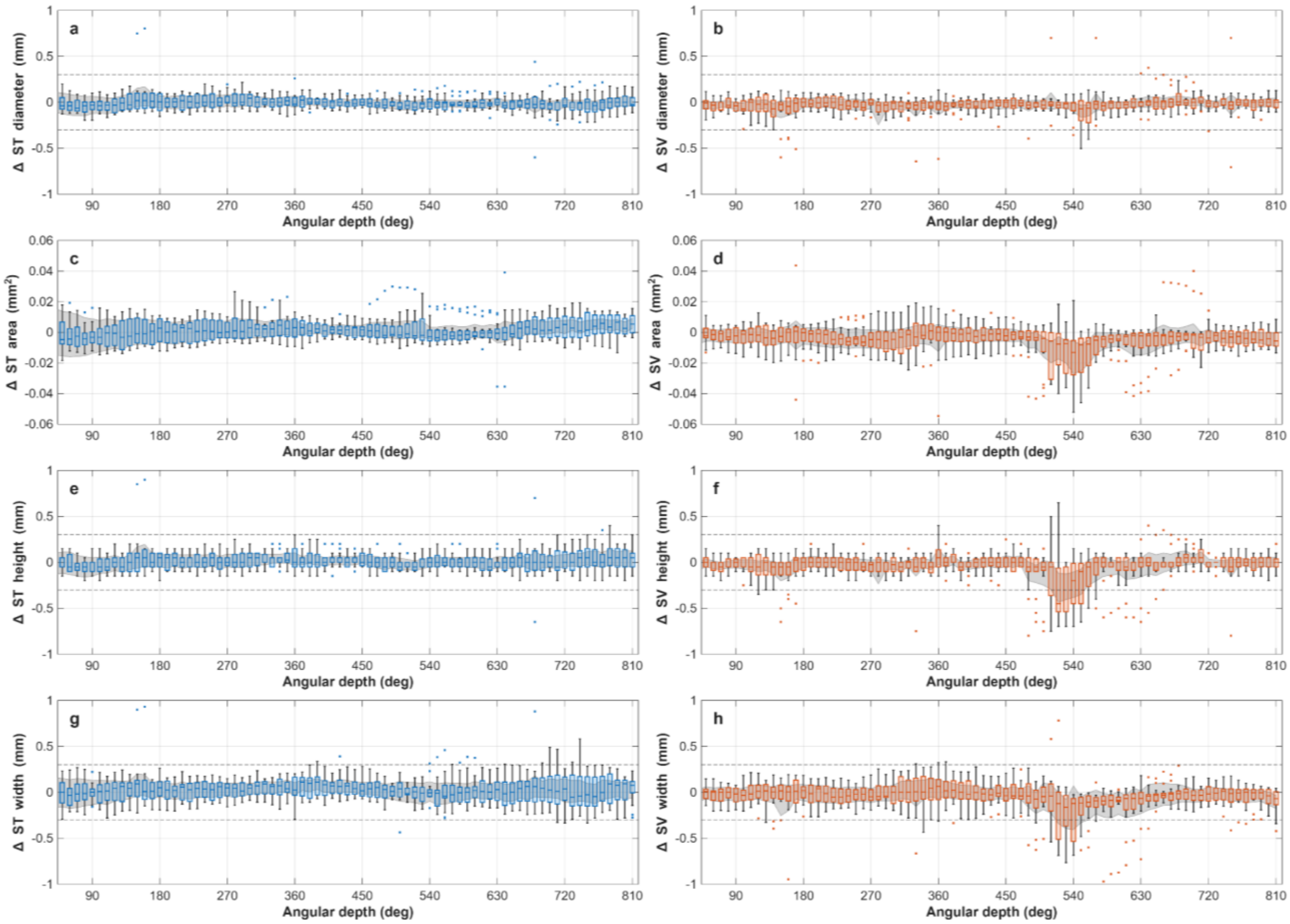

**Supplementary Fig 2. Difference in cross-sectional measurements conducted on the CBCT 300 µm deep learning and SR-PCI segmentations.** Difference (Δ) between cross-sectional measurements for all 15 test cochleae shown at 10-degree angular increments of scala tympani (ST; blue) and the scala vestibuli (SV; orange), diameter **(a, b)**, area **(c, d)**, height **(e, f)**, and width **(g, h)**. Box plots are shown for each scan type with scatter points overlayed to represent individual samples. The median delta is indicated by the central line in each boxplot, with the lower and upper bounds representing the 25th and 75th percentiles. Whiskers span the most extreme non-outlier observations, and outliers are plotted individually as '.' markers. The grey shaded region indicates the 90% confidence interval and the dashed back lines indicate ± a priori equivalence margin of 1 voxel (0.3 mm, 0.09 $mm^2$) used for equivalence testing assessed using Two One-Sided Tests (TOST). The consistent sub voxel equivalence suggests clinically acceptable equivalent measurements between the deep learning and SR-PCI segmentations cross-sectional measurements.

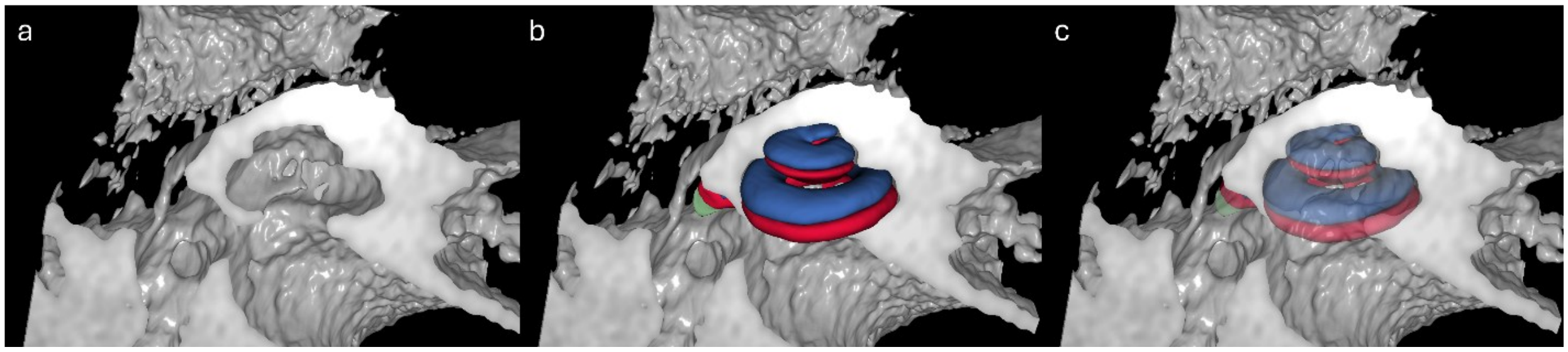

**Supplementary Fig 3. Three-dimensional rendering of a patient scan and the deep learning segmentation.** Clinical patient scan from London Health Sciences Centre, Ontario, Canada showing 3D rendering thresholded to showcase bone encapsulating the cochlea **(a)**, with the opaque **(b)** and transparent segmentation overlayed **(c)**.

**Supplementary Table 1. International clinical computed tomography (CT) dataset acquisition details, including scanner manufacturer, model, and slice thickness of scans from each institution.**

| Institution (Imaging Center) | n | Scanner Manufacturer and Model | Slice Thickness (mm) |
|---|---|---|---|
| Western University (London Health Sciences Centre, Ontario, Canada) | 55 | Xoran<br>• MiniCAT | 0.1 – 0.43 |
| University of North Carolina at Chapel Hill, North Carolina, USA | 46 | Morita<br>• 3D Accuitomo | 0.2 – 0.6 |
| Thomas Jefferson University, Philadelphia, USA | 10 | Siemens<br>• SOMATOM Drive<br>GE<br>• Revolution EVO<br>• Revolution HD<br>• LightSpeed VCT<br>Philips<br>• Ingenuity Core | 0.6 – 1.0 |
| Ludwig-Maximilians-Universität München, Germany | 162 | Siemens<br>• SOMATOM Perspective<br>• SOMATOM go.All<br>• SOMATOM go.Top<br>• SOMATOM go.Up<br>• SOMATOM Definition AS<br>• SOMATOM Definition Edge<br>• SOMATOM Drive<br>• SOMATOM Force<br>• Emotion 16 (2007)<br>GE<br>• Optima CT660<br>• Revolution EVO<br>Philips<br>• Incisive CT | 0.23 – 2.0 |
| | | • iCT 256<br>Toshiba<br>• Aquilion<br>Canon<br>• Aquilion Lightning<br>• Aquilion Prime SP | |
| Comprehensive Hearing Center (University Hospital Würzburg), Würzburg, Germany | 22 | Siemens<br>• NAEOTOM (Alpha photon-counting CT scanner) | 0.2 |
| King Saud University, Riyadh, Saudi Arabia | 228 | Siemens<br>• Volume Zoom<br>• SOMATOM Definition AS<br>• SOMATOM Definition Flash<br>• Somaris 5/ 3D<br>GE<br>• Discovery CT750 HD<br>• Revolution CT<br>• Revolution Maxima<br>• LightSpeed VCT<br>• Optima CT540<br>• SIGNA Artist<br>Philips<br>• iCT 256<br>• Brilliance 64 | 0.23 – 3.0 |